\documentclass[pra,aps,twocolumn,superscriptaddress]{revtex4-2}

\usepackage{graphicx,amsmath,color,soul}

\usepackage[colorlinks,urlcolor=blue,citecolor=blue,linkcolor=blue]{hyperref}

\begin{document}

\title{Bistable soliton switching dynamics in a $\mathcal{PT}$-symmetric coupler with saturable nonlinearity}

\author{Ambaresh Sahoo}
\email{sahoo.ambareshs@gmail.com}
\affiliation{Department of Physics, Indian Institute of Technology Guwahati, Assam 781039, India}

\author{Dipti Kanika Mahato}
\affiliation{Department of Physics, Indian Institute of Technology Guwahati, Assam 781039, India}

\author{A. Govindarajan}
\affiliation{Department of Nonlinear Dynamics, School of Physics,
Bharathidasan University, Tiruchirapalli 620024, India}

\author{Amarendra K. Sarma}
\email{aksarma@iitg.ac.in}
\affiliation{Department of Physics, Indian Institute of Technology Guwahati, Assam 781039, India}

\begin{abstract}
We investigate the switching dynamics in a $\mathcal{PT}$-symmetric fiber coupler composed of a saturable nonlinear material as the core. In such a saturable nonlinear medium, bistable solitons may evolve due to the balance between dispersion and saturable nonlinearity, which we extend in the context of the $\mathcal{PT}$-symmetric coupler. Our investigations of power-controlled and phase-sensitive switching show richer soliton switching dynamics than the currently existing conventional counterparts, which may lead to ultrafast and efficient all-optical switching dynamics at very low power owing to the combined effects of $\mathcal{PT}$ symmetry and saturable nonlinearity. In addition to the input power, the relative phase of the input solitons and saturable coefficient are additional controlling parameters that efficiently tailor the switching dynamics. Also, we provide a suitable range of system and pulse parameters that would be helpful for the practical realization of the coupler to use in all-optical switching devices and photonic circuits. Finally, we develop a variational approach to analytically investigate the switching dynamics in such $\mathcal{PT}$-symmetric couplers that excellently predicts the numerical findings.

\end{abstract}

\maketitle

\section{Introduction}

Parity-time ($\mathcal{PT}$) symmetry, as a concept, got tremendous attention after Bender and Boettcher put forward a very novel perspective on quantum mechanical Hamiltonians \cite{Bender98, Bender99, Bender02}. While experiments were yet to be carried out in the quantum domain, the optical platforms provided support and utility of $\mathcal{PT}$-symmetric ideas \cite{El-Ganainy07, Makris08, Musslimani08}. Following the theoretical predictions, soon there were experimental realizations of the optical $\mathcal{PT}$ symmetry in a linear coupled waveguide system with balanced gain and loss in the two arms of the coupler \cite{Guo09,Ruter10,Regensburger12}. This subsequently led to extending the $\mathcal{PT}$-symmetric theory in other branches of physics quickly \cite{Chang14, Cerjan16, Hang13, Zhang16, Xu16, Xiao21, XuH16, Jing17, Fleury15, Konotop16}. Further progress in understanding the $\mathcal{PT}$ symmetry has been observed once the nonlinearity is included in the system. The interplay between the $\mathcal{PT}$ symmetry and the nonlinearity gave rise to the existence of localized modes \cite{Abdullaev11, Kominis15} and $\mathcal{PT}$-symmetric solitons \cite{Wimmer15, Suchkov16}.

In the context of couplers, which are mostly utilized as all-optical switching devices, when nonlinearity is introduced under the effect of $\mathcal{PT}$ symmetry, several authors reported the existence of the bright, dark, gap, and Bragg solitons, as well as many other interesting phenomena \cite{Suchkov16, Zhiyenbayev19, Miroshnichenko11, Burlak13, Bludov13, Barashenkov13, BludovJO13}. The operation of $\mathcal{PT}$-symmetric couplers, especially with the Kerr nonlinearity, showed improvement as the critical power of switching reduces drastically while maintaining high efficiency \cite{Govindarajan19}. However, a conventional coupler with a Kerr nonlinear medium has a low nonlinear coefficient $n_2$, which requires high input power for switching. To overcome this hindrance, non-Kerr saturable nonlinear media, such as semiconductor doped glass and organic polymers, have been employed due to their higher $n_2$ values compared to pure silica \cite{ Olbright86} and their relatively low saturable intensities \cite{Roussignol87, Coutaz91}. Therefore, for the $\mathcal{PT}$-symmetric couplers, if saturable nonlinearity (SN) is introduced, one can expect to achieve advantageous transmission characteristics over the Kerr one.

Now, one needs to be clear that such media do not support Kerr solitons; instead, there exist bistable solitons, which are basically two solitons having the same pulse width, but different energies and shapes \cite{Kaplan85}. The switching dynamics inside conventional saturable nonlinear couplers (SNCs) by utilizing the bistable solitons has been reported previously \cite{Krolikowski92, Gatz91, Kumar96}, and had been found to be better alternatives for all-optical switching devices \cite{Kumar98}. Furthermore, while there have been several studies demonstrating the existence of stable fundamental soliton, gap soliton, higher-order solitons, and nonlinear modes in different $\mathcal{PT}$-symmetric potentials \cite{Cao14, Zhan16, Zhu16, Li18, Abdullaev20, Moreira20}, there has not been any study on the existence of bistable solitons inside a $\mathcal{PT}$-symmetric SNC and their switching dynamics.

Therefore, in connection with the all-optical switching devices, in this work, we concentrate on the steering dynamics in a $\mathcal{PT}$-symmetric SNC. We first obtain the exact soliton solution which can propagate through such a medium and then observe the transmission characteristics of that pulse for the $\mathcal{PT}$-symmetric coupler and solve the corresponding equation considering the combination of device length and coupling coefficient to be half-beat length. The corresponding theoretical model is described in Sec.\,\ref{Sec2}, along with a discussion on the numerical finding of a soliton solution which can propagate in a saturable nonlinear medium. In Sec.\,\ref{Sec3}, we discuss bistable solitons in the context of a $\mathcal{PT}$-symmetric coupler with SN followed by their power-controlled switching dynamics in Sec.\,\ref{Sec4}. The spatiotemporal characteristics of solitons are illustrated in Sec.\,\ref{Sec5}. In Sec.\,\ref{Sec6}, we discuss the phase-controlled dynamics of the solitons. Section\,\ref{Sec7} is dedicated to the study of Lagrange's variational approach \cite{Bondeson79, Anderson83} to analytically predict the switching dynamics in such a $\mathcal{PT}$-symmetric coupler. The paper is concluded by Sec.\,\ref{Sec8}.

\section{Theoretical Framework} \label{Sec2}

\subsection{Model: $\mathcal{PT}$-symmetric fiber coupler with saturable nonlinearity}
\vspace{-0.2cm}

\begin{figure}[t]
\centering
\begin{center}
\includegraphics[width=0.485\textwidth]{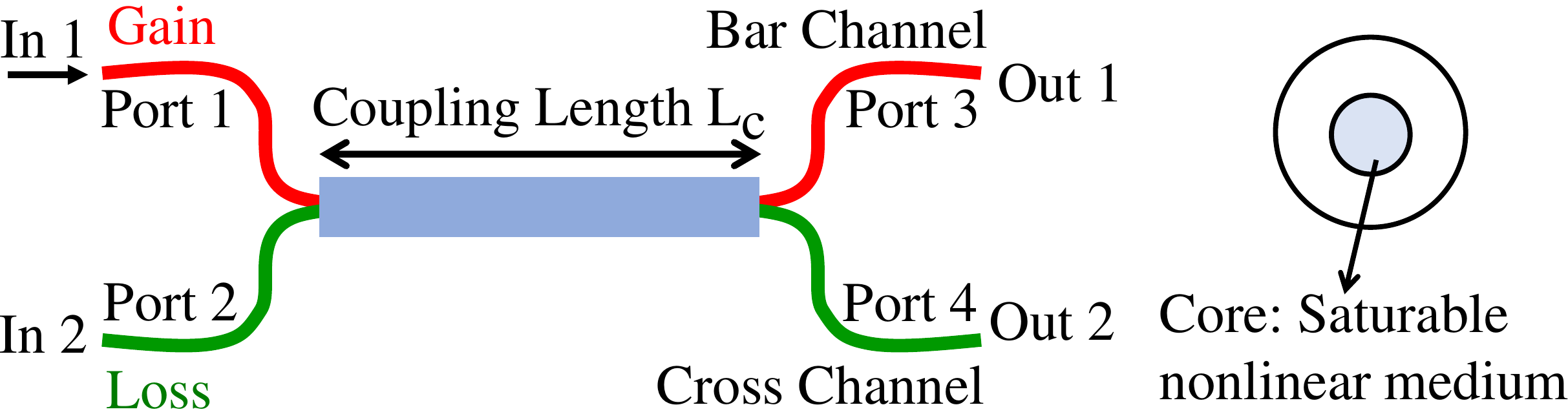}
\caption{(Color online) Schematic diagram of a $\mathcal{PT}$-symmetric directional fiber coupler whose core is made up of a saturable nonlinear medium.}\label{Fig-1}
\end{center} 
\end{figure}
%
We consider a $\mathcal{PT}$-symmetric nonlinear directional coupler with SN (shown in Fig.\,\ref{Fig-1}) in which the medium's nonlinear response saturates beyond a threshold power. The optical pulse propagation in such a coupler can be modeled by a set of coupled-mode nonlinear Schr\"{o}dinger equations (NLSEs) of the slowly varying complex-valued electric field envelopes $A_1(z,t)$ and $A_2(z,t)$ in the two channels \cite{GPAbook2, Govindarajan19}, which can be written in dimensionless form as
\begin{subequations}\label{eq1}
\begin{align}
i\frac{\partial u}{\partial \xi} + \frac{1}{2}\frac{\partial^2 u}{\partial{\tau^2}} +f(|u|^2) u+ \kappa v= i\Gamma u, \label{eq1a} \\
i\frac{\partial v}{\partial \xi} + \frac{1}{2}\frac{\partial^2 v}{\partial{\tau^2}} +f(|v|^2)v+ \kappa u= -i\Gamma v,  \label{eq1b}
\end{align}
\end{subequations}
where $u(\xi,\tau)=A_1/\sqrt{P_0}$ and $v(\xi,\tau)=A_2/\sqrt{P_0}$, with $P_0$ being the peak input power; $\xi=z/L_D$ and $\tau=(t-z/v_g)/t_0$ are respectively the normalized distance and time, with $L_D=t_0^2/|\beta_2(\omega_0)|$, and $t_0$, $v_g$, and $\beta_2$ are the input pulse duration, group velocity of the pulse, and the group-velocity dispersion parameter at the carrier frequency $\omega_0$. The linear coupling coefficient ($K$) and the balanced linear gain/loss coefficient ($G$) are rescaled as $\kappa=K L_D$ and $\Gamma=G L_D$. Also, in our model, we adopt the most used mathematical model of the saturable nonlinear response \cite{Gatz91, Hickmann93}, the dimensionless form of which is represented as
\begin{align}
f(|u|^2)= \frac{|u|^2}{1+s|u|^2},  \label{eq2}
\end{align}
where $s={P_0}/{I_{\rm sat}}$ is the dimensionless refractive index saturation parameter (also known as the strength of saturation), with $I_{\rm sat}$ being the characteristic saturable intensity of the medium. Note that, when $s=0$, Eq.\,(\ref{eq2}) reduces to the well-known Kerr nonlinearity, and when one increases $s$ the saturation effect of the refractive index increases accordingly. 

The system described by Eq.\,(\ref{eq1}) is a $\mathcal{PT}$-symmetric system, whose operational domains are categorized into three regions: unbroken regime ($\kappa > \Gamma$), broken regime ($\kappa < \Gamma$), and an exceptional point ($\kappa = \Gamma$). 
In the unbroken $\mathcal{PT}$-symmetric regime, it has been observed in previous works \cite{Govindarajan19, Sahoo21} that the poor transmission efficiency and unstable soliton evolution make a half-beat length ($L_c=\pi/2\kappa$) $\mathcal{PT}$-symmetric Kerr coupler an inappropriate choice as compared to two-beat length ($L_c=2\pi/\kappa$) $\mathcal{PT}$-symmetric Kerr coupler that shows enhanced switching efficiency. In the case of a $\mathcal{PT}$-symmetric SNC, however, we consider the device to be operated in the unbroken $\mathcal{PT}$ regime with $L_c \kappa=\pi/2$ equal to the half-beat length, which provides better switching dynamics with a lower critical power than the Kerr counterpart.
%
\begin{figure}[t]
\centering
\begin{center}
\includegraphics[width=0.49\textwidth]{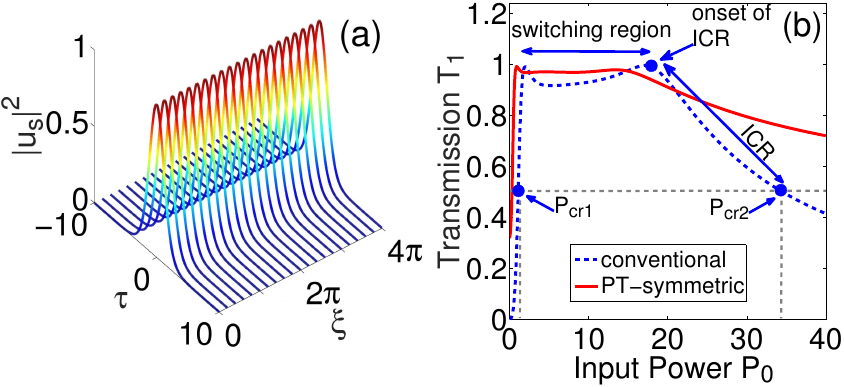}
\caption{(Color online) (a) Soliton solution inside a saturable nonlinear medium, obtained for $ s=1 $ considering the single NLSE. (b) Transmission in the first channel of the conventional ($\Gamma=0$) and $\mathcal{PT}$-symmetric ($\Gamma=0.06$) SNCs with $\kappa=0.1$ and $s=1$, where different operational regions are identified.}\label{Fig-2}
\end{center}
\end{figure}
%
 
\vspace{-0.2cm}
\subsection{Soliton evolution and switching dynamics}
\vspace{-0.3cm}

It is well known that the fundamental Kerr soliton solution, $u={\rm sech} \,(\tau)$, does not satisfy the NLSE with SN that leads to unstable and chaotic evolution dynamics \cite{Gatz91}. Here, we first demonstrate the exact soliton solution corresponding to the single NLSE with SN and use the same soliton solution in the coupled NLSEs  (\ref{eq1a}) and (\ref{eq1b}) to investigate the switching dynamics in a $\mathcal{PT}$-symmetric saturable medium accurately.
For this, we start with a single NLSE with SN \cite{Gatz91}, whose ansatz function $u_s$ takes the following analytical form:
\begin{align} \label{eq3}
u_s(\xi,\tau)=\sqrt{\psi(\tau)}e^{i \beta \xi},
\end{align}
where $\beta$ is the nonlinear propagation constant shift. Inserting this ansatz function [Eq.\,(\ref{eq3})] into the single NLSE and solving the parameters of the ansatz semi-analytically, we obtain the intensity profile of the exact soliton solution ($|u_s|^2$) for the NLSE with SN with strength $s=1$ shown in Fig.\,\ref{Fig-2}(a).
%
\begin{figure}[t]
\centering
\begin{center}
\includegraphics[width=0.495\textwidth]{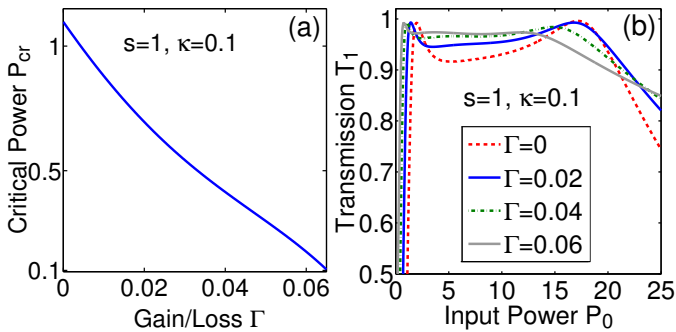}
\caption{(Color online) (a) Relation between critical power $P_{cr}$ and gain/loss parameter $\Gamma$. (b) Effect of gain/loss on the switching dynamics of  both conventional ($\Gamma=0$) and $\mathcal{PT}$-symmetric ($\Gamma\neq0$) SNCs.}\label{Fig-3}
\end{center}
\end{figure}
%
We then consider this soliton solution as the seed solution and solve the set of Eqs.\,(\ref{eq1a}) and (\ref{eq1b}) numerically by applying the symmetrized split-step Fourier method complemented with the fourth-order Runge-Kutta algorithm. Next, to investigate the switching dynamics, we calculate the transmission coefficient $T_j$, which represents the fractional output power in the $j$th channel after propagation of $L_c\kappa$ that is equal to the half-beat length, as
\begin{equation}\label{eq4}
T_j=\frac{\int_{-\infty}^{\infty}|u_{j=1},v_{j=2}(L_c,\tau)|^2 d\tau}{\int_{-\infty}^{\infty}(|u(L_c,\tau)|^2 + |v(L_c,\tau)|^2)d\tau}=\frac{P_j}{P_1+P_2},
\end{equation} 
where $P_1=\int_{-\infty}^{\infty}|u(L_c,\tau)|^2 d\tau$ and $P_2=\int_{-\infty}^{\infty}|v(L_c,\tau)|^2 d\tau$ are the output powers of the transmitted pulse in the output ports of the two channels. In order to investigate the switching dynamics, we plot the transmission coefficient $T_1$ [evaluating Eq.\,\eqref{eq4}] in the first channel of  conventional (blue dashed curve) and $\mathcal{PT}$-symmetric (red solid line) SNCs for coupling coefficient $\kappa=0.1$ and coupling length $L_c=\pi/2\kappa=5\pi$ in Fig. \ref{Fig-2}(b). The various regions of operation for all-optical switching for the conventional coupler are illustrated in the same figure, which can also be translated to the $\mathcal{PT}$-symmetric one. Here, the switching curve shows four basic regimes: the coupling region (below the lower-branch critical power $P_{cr1}$); the switching region [between $P_{cr1}$ and the onset of the intermediate coupling region (ICR)]; the ICR; and another coupling region above the higher-branch critical power $P_{cr2}$, after which the power remains in the second channel for any further increase in the input power $P_0$. Thus, for efficient switching, one must concentrate solely on the range of input power corresponding to the switching region. Furthermore, we find that the higher branch of critical power $P_{cr2}$ appears to be too high to be useful for any optical switching. Therefore, we focus on the lower-branch critical power $P_{cr1}$ and denote it as $P_{cr}$ for the rest of our work in order to operate the SNC as an efficient all-optical switching device.
Now, with the introduction of balanced gain/loss, the critical power $P_{cr}$ is reduced considerably, and both the switching steepness and transmission efficiency are enhanced. For further investigation of how the gain/loss parameter $\Gamma$ affects the switching dynamics, we plot $P_{cr}$ as a function of $\Gamma$ for specific values of $\kappa=0.1$ and $s=1$ in Fig.\,\ref{Fig-3}(a). This plot indicates that when $\Gamma$ approaches the singularity (i.e., $\Gamma=\kappa=0.1$ in this particular case), no switching occurs, implying that, regardless of the input power, all of the power remains in the launching channel. The corresponding transmission is shown in Fig.\,\ref{Fig-3}(b) for the four different values of $\Gamma=0,\,0.02,\,0.04$, and $0.06$. As we gradually increase $\Gamma$ from $0$ to $0.06$, we observe a lower critical power with better transmission efficiency. Hence, based on these results, we fix the gain/loss value of the $\mathcal{PT}$-symmetric SNC to be $\Gamma=0.05$ for further investigations on the switching dynamics.

\vspace{-0.2cm}

\section{Bistable Soliton dynamics in a $\mathcal{PT}$-symmetric saturable coupler}    \label{Sec3}
\vspace{-0.3cm}
%
\begin{figure}[t]
\centering
\begin{center}
\includegraphics[width =0.49\textwidth]{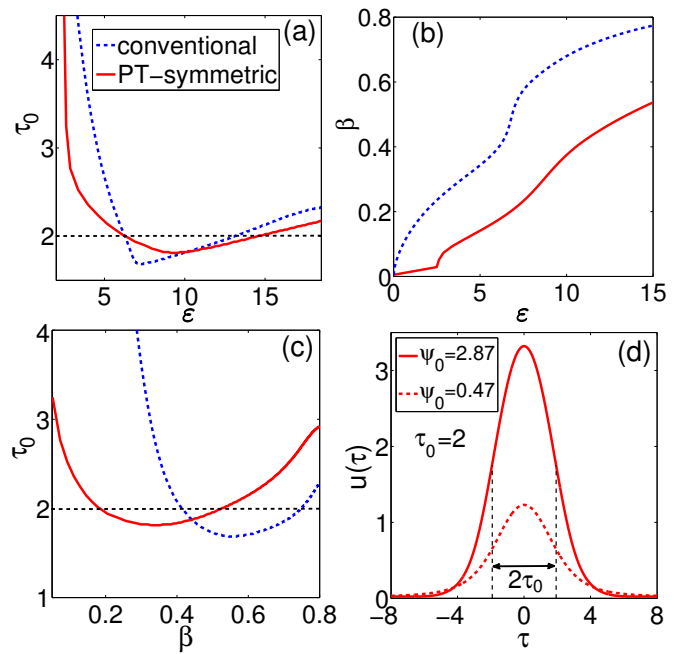}
\caption{(Color online) Variation of output pulse energy $\varepsilon$ of the coupler as a function of (a) output pulse width $\tau_0$ and (b) propagation constant shift $\beta$ of the input seed soliton [Eq.\,\eqref{eq3}]. (c) The variation of $\tau_0$ as a function of $\beta$. Solid red curves in (a)-(c) represent the case with $\mathcal{PT}$ symmetry, whereas dashed blue curves represent the conventional one.
(d) Amplitude $u(\tau)$ of the bistable solitons at the output of the $\mathcal{PT}$-symmetric SNC corresponding to $\tau_0=2$ for two energies $\varepsilon=6.22$ and $\varepsilon=14.65$, respectively. The horizontal dashed lines in (a) and (c) and the vertical dashed lines in (d) indicate that $\tau_0=2$.}
\label{Fig-4}
\end{center}
\end{figure}

Before going deeper into the details of soliton switching dynamics in $\mathcal{PT}$-symmetric SNCs, we first investigate the properties of bistable solitons and their dynamics, which occur inherently in saturable nonlinear systems. Previous researches have looked into the fundamental properties of bistable solitons in saturable nonlinear media, where there exist two solitons with the same pulse width but different shapes and energies \cite{Kaplan85, Gatz91}. In our work, it is interesting to investigate how the seed soliton [Eq.\,\eqref{eq3}], which is inherently bistable in nature for a specific range of ansatz parameters, evolves in the couplers and retains their properties as well as individual switching characteristics. 
%
\begin{figure}[t]
\centering
\begin{center}
\includegraphics[width =0.485\textwidth]{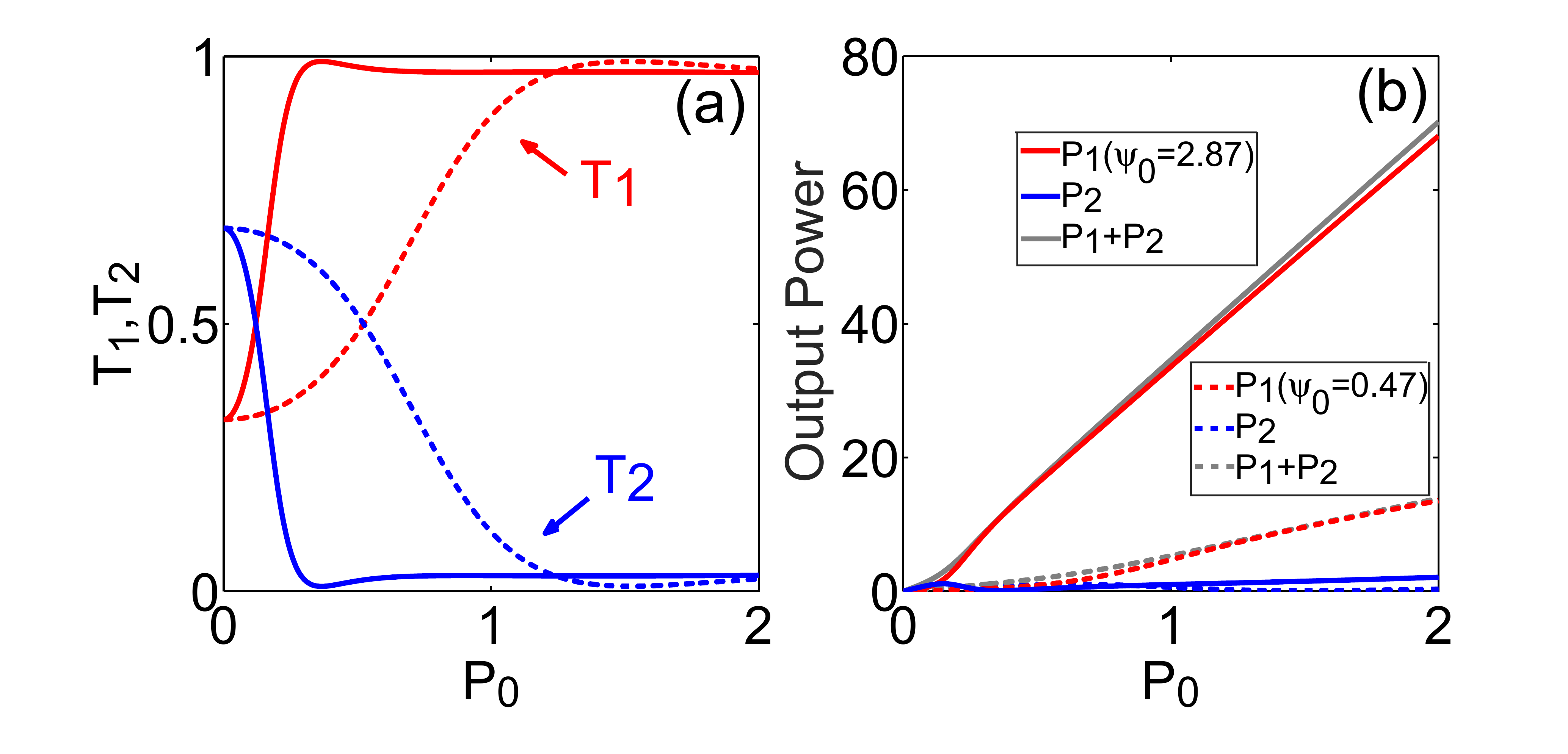}
\caption{(Color online) (a),(b) Switching dynamics of bistable solitons in the $\mathcal{PT}$-symmetric SNC corresponding to the pulse width $\tau_0=2$ and two input peak amplitudes $\psi_0=0.47$ (dashed curves) and $\psi_0=2.87$ (solid curves).} \label{Fig-5}
\end{center}
\end{figure}
%
To investigate the properties of bistable solitons, we first define the dimensionless soliton energy $\varepsilon$  as
\begin{equation} \label{eq5}
\varepsilon=\int_{-\infty}^{\infty}|u(\tau)|^2 d\tau.
\end{equation}
Next, for a given value of soliton peak amplitude $\psi(\tau=0)=\psi_0$, we compute the output pulse width $\tau_0$ and the propagation constant shift $\beta$ [presented in Eq. (\ref{eq3})]. In Fig.\,\ref{Fig-4}(a), we plot the variation of pulse energy $\varepsilon$ [Eq.\,\eqref{eq5}] at the first output port of the coupler with $L_c\kappa=\pi/2$ as a function of the output pulse width $\tau_0$ by varying the peak amplitude $\psi_0$ of the input seed soliton. Similarly, we plot the variation of $\varepsilon$ as a function of $\beta$ [see Fig.\,\ref{Fig-4}(b)] and $\tau_0$ vs $\beta$ [see Fig.\,\ref{Fig-4}(c)] for the system parameters $\kappa=0.1$,  $\Gamma=0.05$, and $s=1$. The bistability nature is clearly evident in both the conventional (dashed blue curves) and $\mathcal{PT}$-symmetric (solid red curves) SNCs [there exist two values of $\varepsilon$ and $\beta$ at the same $\tau_0=2$, as shown by the horizontal dashed line in Figs.\,\ref{Fig-4}(a) and \ref{Fig-4}(c)]. Furthermore, these plots indicate that the addition of a $\mathcal{PT}$-symmetric potential modifies the parameter space of the soliton in comparison to the conventional SNC. The bistable solitons inside the $\mathcal{PT}$-symmetric SNC with two different energies $\varepsilon=6.22$ and $\varepsilon=14.65$ corresponding to the pulse width $\tau_0=2$ are shown in Fig.\,\ref{Fig-4}(d). Here, the higher amplitude soliton (solid red curve) corresponds to the higher input peak amplitude $\psi_0=2.87$, while the lower amplitude soliton (dashed red curve) corresponds to the lower input peak amplitude $\psi_0=0.47$. Thus, by examining Figs.\,\ref{Fig-4}(a)-\ref{Fig-4}(d), we can conclude that the bistable solitons exist in the unbroken regime of a $\mathcal{PT}$-symmetric SNC in the weak coupling regime.

Next, the switching dynamics of the individual bistable solitons obtained in Fig.\,\ref{Fig-4}(d) are investigated in Figs.\,\ref{Fig-5}(a) and \ref{Fig-5}(b). Here, we observe two independent steering dynamics for two solitons depending on the soliton amplitude. The critical power of switching corresponding to the pulse with $\psi_0=2.87$ is lower than that of the pulse with $\psi_0=0.47$. This is because the nonlinear response of the medium becomes saturated earlier in the case of the pulse with high input peak amplitude, causing the switching to occur faster than in the case of the pulse with low input peak amplitude.
In the following sections, our investigations mainly focus on the switching dynamics of a single soliton case for a given pulse width, which have applications in all-optical switching devices and photonic circuits.

\section{Power-controlled switching dynamics}   \label{Sec4}

\begin{figure}[t]
\centering
\begin{center}
\includegraphics[width=0.495\textwidth]{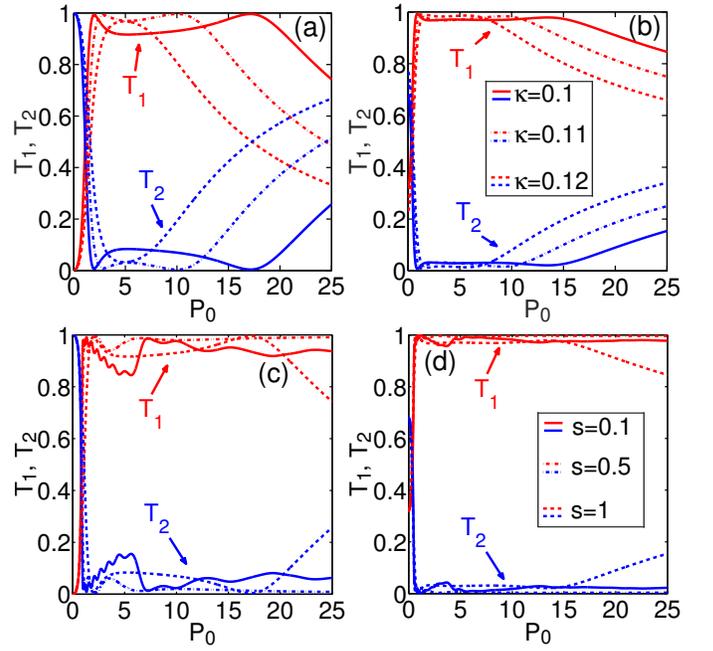}
\caption{(Color online) Switching dynamics of (a) conventional and (b) $\mathcal{PT}$-symmetric ($\Gamma=0.05$) SNCs for different $\kappa$ with fixed $s=1$. Panels (c) and (d) represent the same as (a) and (b) but for different $s$ with fixed $\kappa=0.1$.}\label{Fig-6}
\end{center}
\end{figure}

To discuss the power-controlled switching dynamics, we assume that in the presence of a coupling coefficient ($\kappa\neq0 $), the soliton is always launched into the input port (port 1) of the first channel, and the second channel is kept empty so that
\begin{align} \label{eq6}
u(\xi=0,\tau)  =u_s(\xi=0,\tau) ~~ {\rm and}	~~
v(\xi=0,\tau)  = 0.
\end{align}
With the initial conditions above in Eq.\ (\ref{eq6}), we discuss the effect of coupling coefficient $\kappa$ on the switching dynamics for  conventional ($\Gamma=0$) and  $\mathcal{PT}$-symmetric ($\Gamma=0.05$) SNCs by setting the saturation to be maximum, $s=1$, shown in Figs. \ref{Fig-6}(a) and \ref{Fig-6}(b). For both the SNCs, the lower value of $\kappa$ shows lower critical power and better transmission efficiency across a wider input power range.
In our analysis, we find that the region of the power-controlled stable soliton switch between two channels of the coupler occurs for relatively lower coupling coefficients, beyond which the power remains in the launching core. This lower coupling coefficient results in a very low $\Gamma$ for the power-controlled stable soliton switch with a very low critical power, as shown in Fig.\,\ref{Fig-3}, which is inherent in a half-beat length saturable $\mathcal{PT}$-symmetric coupler operating in the unbroken regime [$\Gamma(=0.05)<\kappa(=0.1)$].
%
%
%
\begin{figure}[t]
\centering
\begin{center}
\includegraphics[width=0.485\textwidth]{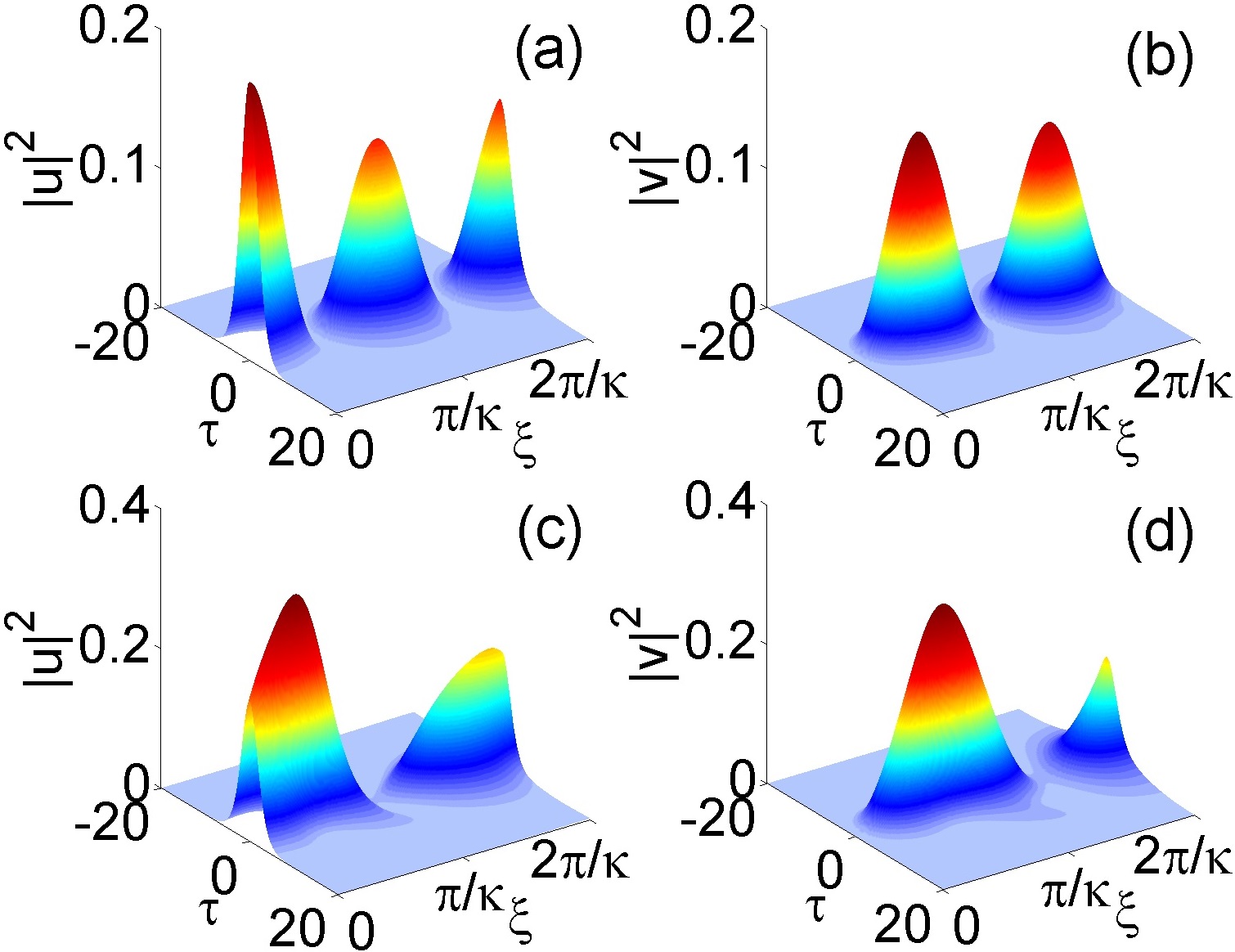}
\caption{(Color online) Spatiotemporal evolution of solitons in (a),(b) a conventional SNC ($\Gamma=0$) and (c),(d) a $\mathcal{PT}$-symmetric SNC ($\Gamma=0.05$) with input power $P_0=0.2$. The system parameters are taken to be $\kappa=0.1$ and $s=1$.}\label{Fig-7}
\end{center}
\end{figure}
%
%
\begin{table}[b] 
\caption{Normalized values of critical power $P_{cr}$ for different strengths of saturation and coupling coefficients.} \label{Table1}
\begin{center}
\begingroup
\setlength{\tabcolsep}{6pt} 
\renewcommand{\arraystretch}{1.5}
\begin{tabular}{ c c c c } 
\hline
\hline
Fixed & Varying & $P_{cr,{\rm conventional}}$ & $P_{cr,\mathcal{PT}}$\\
parameter & parameter &  ($\Gamma=0$) & ($\Gamma=0.05$)\\
\hline
& $\kappa=0.1$  & 1.10 & 0.30 \\ 
$s=1$ & $\kappa=0.11$ & 1.30 & 0.38 \\ 
& $\kappa=0.12$ & 1.59 & 0.47 \\
\hline
& $s=0.1$ & 0.82 & 0.37 \\ 
$\kappa=0.1$  & $s=0.5$ & 0.86 & 0.32 \\ 
& $s=1$ & 1.10 & 0.30 \\ 
\hline
\hline
\end{tabular}
\endgroup
\end{center} \vspace{-0.3cm}
\end{table}

%
\begin{figure}[t]
\centering
\begin{center}
\includegraphics[width=0.485\textwidth]{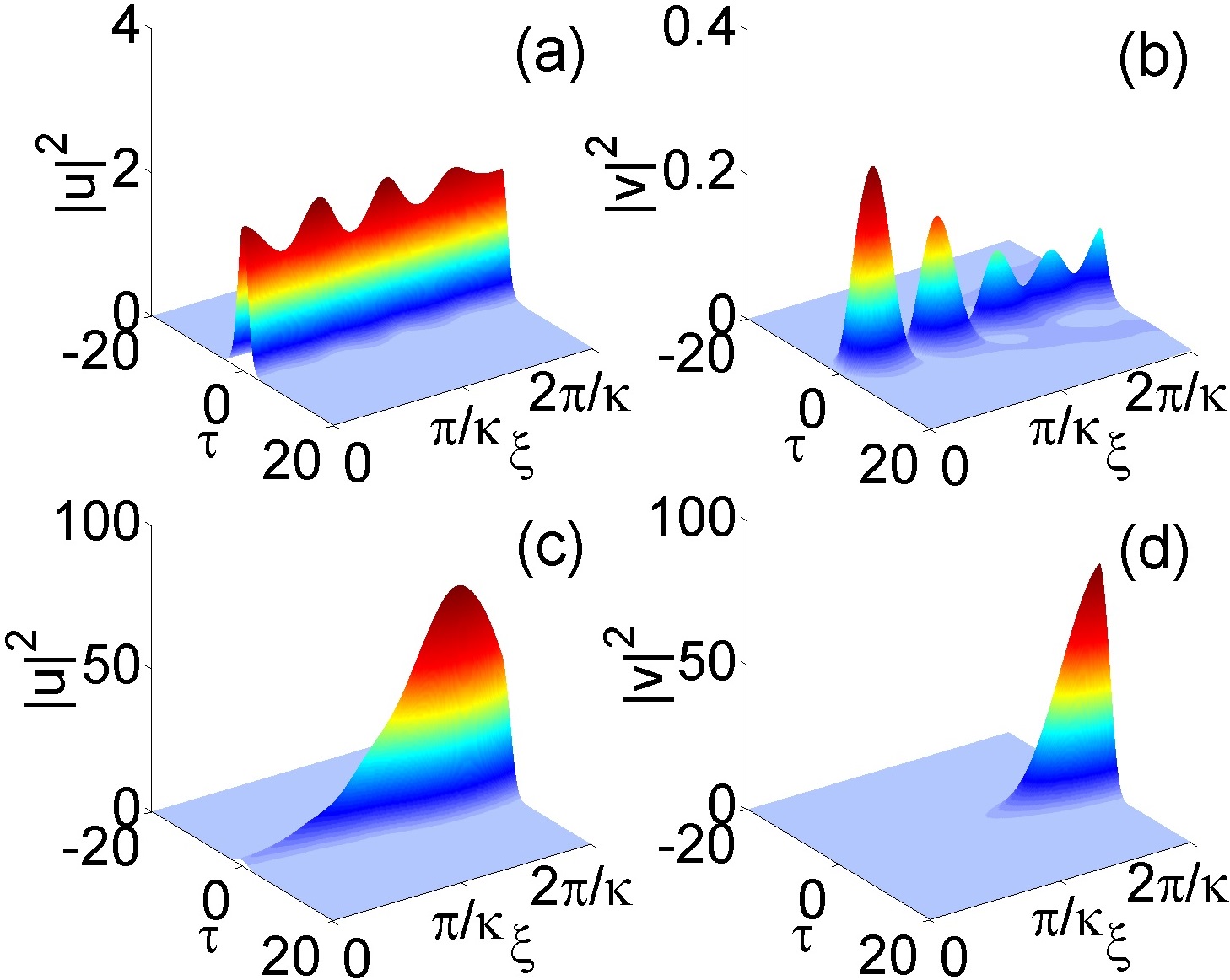}
\caption{ (Color online) Spatiotemporal evolution of solitons in (a),(b) a conventional SNC, and (c),(d) a $\mathcal{PT}$-symmetric SNC with input power $P_0=2$. All the other parameters are the same as in Fig.\,\ref{Fig-7}} \label{Fig-8}
\end{center}
\end{figure}

Next, we study the role of the strength of SN, $s$, for the same two couplers in the weaker coupling regime by setting $\kappa=0.1$, shown in Figs. \ref{Fig-6}(c) and \ref{Fig-6}(d). In both the conventional and the $\mathcal{PT}$-symmetric SNCs, we observe a Kerr coupler-like behavior for lower saturation strengths, $s=0.1$ and $s=0.5$. In contrast, the typical effect of SN in the high input power domain is observed once the couplers are operated with maximum saturation $s=1$. In Table I, we provide a list of the the normalized values of critical powers for the two types of SNCs, showing the variations of $\kappa$ and $s$. We observe that,  when the strength of SN increases gradually, unlike a conventional SNC, the critical power of switching tends to decrease for a $\mathcal{PT}$-symmetric one.
Thus, based on the values of critical powers $P_{cr}$ given in Table \ref{Table1}, we can conclude that, for the weaker coupling region ($\kappa=0.1$) and maximum saturation strength ($s=1$), a $\mathcal{PT}$-symmetric coupler is more efficient compared to the conventional one.

\vspace{-0.2cm}

\section{Spatiotemporal soliton Dynamics}   \label{Sec5}
\vspace{-0.3cm}

In this section, we illustrate the spatiotemporal evolution dynamics of a soliton inside the two cores of the two SNCs. In Figs. \ref{Fig-7}(a)-\ref{Fig-7}(d), with the previously chosen values of $\kappa=0.1$, $s=1$ and $\Gamma=0.05$, we plot the corresponding evolution dynamics inside the couplers in the linear regime where the input power is chosen to be $P_0=0.2$, lower than the critical power $P_{cr}=0.3$. One can predict these evolution trends inside the conventional and $\mathcal{PT}$-symmetric SNCs by following the switching curves in Fig. \ref{Fig-2}(b). In this linear regime, the soliton couples back and forth inside the two channels and eventually exits from the output port of the first channel. On the other hand, if a soliton is launched in the nonlinear regime, where the input power is set to be $P_0=2$, higher than $P_{cr}$, we observe that the soliton switches back to the launching core and remains in there in the case of a conventional SNC [see Figs. \ref{Fig-8}(a) and \ref{Fig-8}(b)]. However, for the case of a $\mathcal{PT}$-symmetric SNC, we observe that the soliton power comes out of the second channel as shown in Figs. \ref{Fig-8}(c) and \ref{Fig-8}(d). Thus, with the inclusion of gain/loss in the SNC, the nonlinear switching takes a different route than the conventional one.

\vspace{-0.0cm}

\section{Switching dynamics by controlling the relative phase}   \label{Sec6}
\vspace{-0.2cm}

\begin{figure}[t]
\centering
\begin{center}
\includegraphics[width=0.49\textwidth]{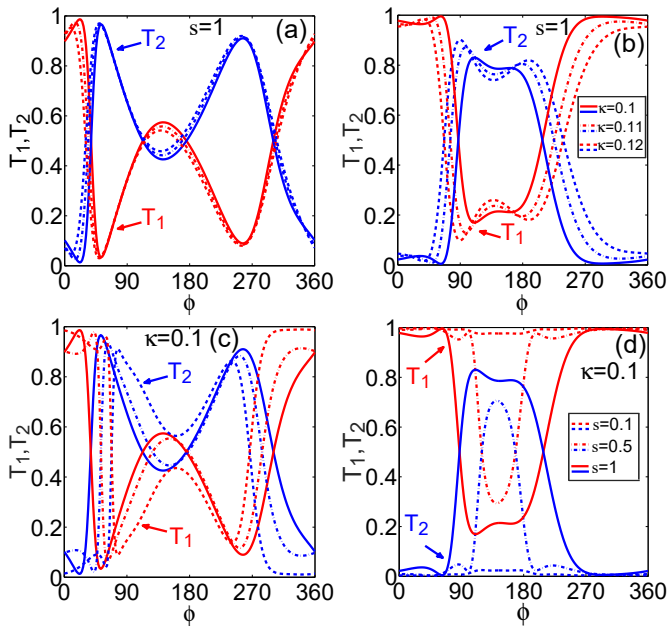}
\caption{(Color online) Phase-controlled switching dynamics of (a) conventional and (b) $\mathcal{PT}$-symmetric ($\Gamma=0.05$) SNCs for different $\kappa$ with fixed $s=1$. Panels (c),(d) represent the same as (a),(b) but for different $s$ with fixed $\kappa=0.1$.}\label{Fig-9}
\end{center}
\end{figure}
%

In addition to the power-controlled switching, for the completeness of the study, we investigate phase-sensitive switching dynamics as another type of all-optical switching device in this section. The basic idea of a phase-sensitive switching is that, by adjusting the input power and the relative phase of a weaker pulse in the second channel, one can control the steering dynamics of the stronger pulse which is launched into the first channel \cite{Trillo91, Romagnoli92}. Therefore, we consider two soliton pulses launched into the input ports (port 1 and port 2) of the two channels as
%
\begin{align} \label{eq7}
u(\xi=0,\tau) =u_s  ~~ {\rm and}	~~
v(\xi=0,\tau)  = \frac{u_s}{\sqrt{r}}e^{i\phi}, 
\end{align}
%
where $r$ is the power ratio factor and $\phi$ is the relative phase between the two soliton pulses. For a typical value $r=4$, and considering Eq.\,(\ref{eq7}) as inputs, we show the effect of coupling coefficient for both conventional and $\mathcal{PT}$-symmetric SNCs considering the same set of $\kappa$ as discussed in the power-controlled case. The $\mathcal{PT}$-symmetric SNC displays considerably better sharp phase-sensitive switching as compared to the conventional one, with more than 90$\%$ transmittance over certain ranges of $\phi$ [see Figs. \ref{Fig-9}(a) and \ref{Fig-9}(b)]. We observe that the lower the coupling coefficient is, the higher the transmission efficiency is for a certain range of $\phi$.
We also plot the effect of the strength of SN on the phase-controlled switching dynamics while considering the previous set of $s$ values as in the power-controlled case. Here, we observe that, for the conventional SNC, the lower $s$ value ($s=0.1$) displays higher transmission efficiency over certain range of $\phi$ [see Fig. \ref{Fig-9}(c)], whereas in the $\mathcal{PT}$-symmetric SNC, for the lower strength of saturation, the transmittance in the first channel is more than 97$\%$ for any $\phi$ value [dashed red curve Fig. \ref{Fig-9}(d)]. Thus, by controlling the phase of the weaker pulse, we can steer the stronger pulse to come out from the desired channel. By observing these results, we can conclude again that a $\mathcal{PT}$-symmetric SNC is better than a conventional SNC for use as a phase-controlled switching device as well.

\vspace{-0.2cm}
\section{Variational approach to study the switching dynamics}   
\label{Sec7}
\vspace{-0.2cm}

To investigate the switching dynamics in such $\mathcal{PT}$-symmetric couplers analytically, we adopt Lagrange's variational technique \cite{Bondeson79, Anderson83}. This technique has proved to be an effective analytical tool for describing pulse dynamics in both conservative and various dissipative systems \cite{GPAbook1, Kaup95, Cerda98, Sahoo17,SahooAM19, Sahoo19, SahooAKS21}. The variational method has also been applied to conventional couplers \cite{Ankiewicz95,  Kumar04} and systems with complex potentials \cite{Barashenkov14, Hu17}. The success of the variational method relies on the proper choice of the {\it Ansatz} function, which in our case yields excellent predictions of the switching dynamics in $\mathcal{PT}$-symmetric couplers.

To apply the variational method, we first write the coupled-mode equations [Eqs. (1a) and (1b)] in the form of perturbed coupled NLSEs,
\begin{align} \label{eq8}
i{\partial_\xi \psi_{1,2}} + \frac{1}{2}{\partial_\tau^2 \psi_{1,2}} +f\left(|\psi_{1,2}|^2\right)\psi_{1,2} + \kappa \psi_{2,1}=i\epsilon_{1,2},
\end{align}
where $\psi_{1}(\xi,\tau)=u$, $\psi_{2}(\xi,\tau)=v$, and the nonconservative terms (gain and loss) are included through the perturbation terms $\epsilon_1=\Gamma \,\psi_{1}$ and $\epsilon_2=-\Gamma \,\psi_{2}$. Now, introducing a Lagrangian density $\mathcal{L_D}=\sum_{j=1,2}\{\frac{i}{2}(\psi_j^*\partial_\xi\psi_j -\psi_j\partial_\xi\psi_j^* )-\frac{1}{2}|\partial_\tau \psi_j|^2 +\frac{1}{s}(|\psi_j|^2 - \frac{1}{s}\ln[1+s|\psi_j|^2 ] ) -2{\rm Re}[i\epsilon_j\psi_j^*]  \} +2\kappa\,{\rm Re}[\psi_1^*\psi_2]$ appropriate for Eq.\,\eqref{eq8} and considering a simple mathematical form of the {\it Ans\"{a}tze} $\psi_{1,2}(\xi,\tau)=\left[{E_{1,2}(\xi)\,\eta(\xi)}/{2}\right]^{1/2}{\rm sech}[\eta(\xi)\,\tau] \,\exp[i\phi_{1,2}(\xi)]$, we obtain a reduced Lagrangian $\left(L = \int_{-\infty}^{\infty}\mathcal{L_D}\, d\tau \right)$:\\
\begin{align} \label{eq9}
L &= \sum_{j=1,2}\left\{- ({\partial_\xi\phi_j}) E_j -\frac{1}{6}\eta^2 E_j + \frac{1}{s}\left(E_j - \frac{1}{2s\eta}\mathcal{F}_j^2 \right)   \right. \nonumber \\ &\hspace{0.8cm}\left. -2\,{\rm Re}\int_{-\infty}^{\infty}\left[i\epsilon_j \psi_{j}^* \right]d\tau \right\} +2\kappa\sqrt{E_1 E_2}\, \cos\Phi. 
\end{align}
Here $\mathcal{F}_j={\rm acosh}(1+s\eta E_{j})$, $\Phi=\phi_1 - \phi_2$, and the five {\it Ans\"{a}tze} parameters $E_{1,2}$ (pulse energy), $\eta$ (inverse of temporal pulse width), and $\phi_{1,2}$ (phase) are  assumed to evolve with $\xi$. 
The next step is to use Euler-Lagrange equation for each ansatz parameter to obtain a set of coupled ordinary differential equations (ODEs) describing overall spatiotemporal soliton dynamics, which results in the following set of coupled ODEs and one self-consistent equation:
%
\begin{figure}[t]
\centering
\begin{center}
\includegraphics[width=0.49\textwidth]{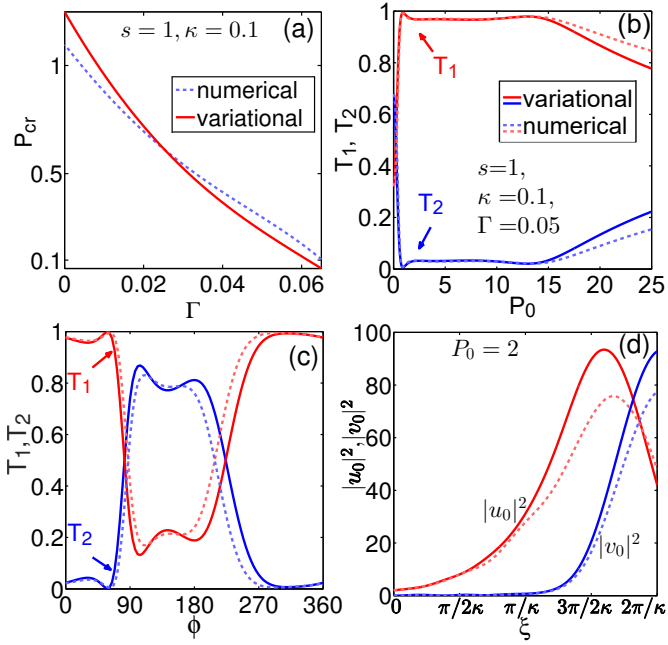}
\caption{(Color online) (a) Critical power vs gain/loss parameter from variational approach and from numerical data [Fig.\,\ref{Fig-3}(a)]. (b) Power-controlled switching, (c) phase controlled switching, and (d) evolution of peak intensities in two channels, with  respective numerical data are compared from Figs.\,\ref{Fig-6}(b), \ref{Fig-9}(d), and \ref{Fig-8}(c,d). Here, $P_0={E_1}(\xi=0)/2$, $|u_0|^2=E_1 \,\eta/2$, $|v_0|^2=E_2 \,\eta/2$, $\phi=-\Phi$. The coupled equations [Eqs.\,\eqref{eq10}-\eqref{eq13}] are solved with initial conditions (at $\xi=0$): $E_2=10^{-6}$, $\Phi=0$, $\eta=1$ for (a) and (b) with varying $E_1$; $E_1=2$, $E_2=1/2$, $\eta=1$ for (c) with varying $\Phi$; $E_1=4$, $E_2=10^{-6}$, $\Phi=0$, $\eta=1$ for (d). }\label{Fig-10}
\end{center} \vspace{-0.5cm}
\end{figure}
\begin{align}
\frac{d E_1}{d\xi}&=2\kappa \sqrt{E_1 E_2}\,\sin\Phi +2\,\Gamma \,E_1, \label{eq10} 
\end{align}
\begin{align}
\frac{d E_2}{d\xi}&=-2\kappa \sqrt{E_1 E_2}\,\sin\Phi -2\,\Gamma\,E_2, \label{eq11} \\
\frac{d \Phi}{d \xi}&=\frac{1}{s}\left(\mathcal{G}_2 -\mathcal{G}_1 \right) +\kappa\frac{\left(E_2 - E_1 \right)}{\sqrt{E_1E_2}} \cos\Phi,    \\
\eta= &\frac{3}{2}\frac{\left(\mathcal{F}_1^2+\mathcal{F}_2^2 \right) - 2s\eta \left( E_1 \mathcal{G}_1 + E_2 \mathcal{G}_2  \right)}{s^2\eta^2(E_1 +E_2)}, \label{eq13}
\end{align}
where $\mathcal{G}_{j}=\mathcal{F}_j/\sqrt{s\eta E_j(2+s \eta E_{j})}$. This set of equations [Eqs.\,\eqref{eq10}-\eqref{eq13}] enable one to get physical insights into the system. For example, the $\Gamma$ term appears in both the energy equations [Eqs.\,\eqref{eq10} and \eqref{eq11}] with $+$ and $-$ signs corresponding to gain and loss, while the coupling and phase terms ($\kappa\sin\Phi$) associated with these two equations imply back and forth energy oscillations between two channels.

We solve the set of coupled equations [Eqs.\,\eqref{eq10}-\eqref{eq13}] semianalytically to evaluate the evolution of individual pulse parameters and investigate the switching dynamics. The results are summarized by solid curves in Figs.\,\ref{Fig-10}(a)-\ref{Fig-10}(d). Here, the numerical findings (light dashed curves) are well supported by the variational predictions (solid curves). However, a slight deviation between these two is observed, which can be attributed to the fact that the single NLSE with SN does not have a closed mathematical form of the {\it Ansatz}. For that reason, we rely on the $sech$ form of the {\it Ansatz} for the variational method.

 \vspace{-0.2cm}
 
\section{Conclusion}    \label{Sec8}

To conclude, we have investigated the steering and switching dynamics of bistable solitons in a $\mathcal{PT}$-symmetric coupler with saturable nonlinearity. One of the objectives of the work was to investigate if the scheme of optical $\mathcal{PT}$ symmetry enhances the transmission characteristics of the saturable coupler compared to the one with Kerr nonlinearity as well as the conventional counterparts. We find the answer is affirmative. The soliton is found to be stable while propagating through the coupler. It turns out that both the power and the phase-controlled mode of switching works well in the $\mathcal{PT}$-symmetric saturable nonlinear coupler.
Also, we adopt Lagrange's variational technique to analytically capture the switching dynamics of such a coupler. The numerical findings are well supported by the variational predictions. It is anticipated that, owing to the huge reduction in the peak power and rich transmission characteristics, the proposed $\mathcal{PT}$-symmetric coupler with saturation nonlinearity would be of great utility for many signal processing applications. Apart from that, there could be a resurgence of research works in this area considering the promising results reported in this work.

\end{document}